\documentclass[sigconf]{acmart}

\usepackage{pifont}
\usepackage{wrapfig} 
\usepackage{amsmath,amsfonts}
\usepackage{array}
\usepackage[caption=false,font=normalsize,labelfont=sf,textfont=sf]{subfig}
\usepackage{textcomp}
\usepackage{stfloats}
\usepackage{url}
\usepackage{verbatim}
\usepackage{graphicx}
\usepackage{paralist}
\usepackage{amsmath,amsfonts}
\usepackage{array}
\usepackage[caption=false,font=normalsize,labelfont=sf,textfont=sf]{subfig}
\usepackage{textcomp}
\usepackage{stfloats}
\usepackage{url}
\usepackage{verbatim}
\usepackage{graphicx}
\usepackage[ruled,vlined,linesnumbered]{algorithm2e}
\usepackage{algpseudocode}
\usepackage{xcolor}
\usepackage{multirow}
\usepackage{amsfonts}
\usepackage{xspace}
\usepackage{colortbl}
\usepackage{makecell}
\usepackage{microtype}
\usepackage{enumitem}
\usepackage{pifont}
\usepackage{booktabs}
\usepackage{bbold}
\usepackage[switch]{lineno}
\usepackage{listings}
\usepackage{xcolor}
\usepackage{tcolorbox}
\newcommand{\tool}{\textit{Drivora}\xspace}

\AtBeginDocument{%
  }

\copyrightyear{2026}
\acmYear{2026}
\setcopyright{cc}
\setcctype{by}
\acmConference[ICSE-Companion '26]{2026 IEEE/ACM 48th International Conference on Software Engineering}{April 12--18, 2026}{Rio de Janeiro, Brazil}
\acmBooktitle{2026 IEEE/ACM 48th International Conference on Software Engineering (ICSE-Companion '26), April 12--18, 2026, Rio de Janeiro, Brazil}
\acmPrice{}
\acmDOI{10.1145/3774748.3787599}
\acmISBN{979-8-4007-2296-7/2026/04}

\acmConference[ICSE 2026]{48th International Conference on Software Engineering}{April 2026}{Rio de Janeiro, Brazil}
\acmISBN{978-1-4503-XXXX-X/2018/06}

\begin{document}

\title{\tool: A Unified and Extensible Infrastructure for Search-based Autonomous Driving Testing}

\author{Mingfei Cheng}
\email{mfcheng.2022@smu.edu.sg}
\orcid{0000-0002-8982-1483}
\affiliation{%
  \institution{Singapore Management University, Singapore}
  \country{}
}


\author{Lionel Briand}
\email{lbriand@uottawa.ca}
\orcid{0000-0002-1393-1010}
\affiliation{%
 \institution{University of Ottawa, Canada}
 \institution{Lero Centre, University of Limerick, Ireland}
 \country{}
 }

\author{Yuan Zhou}
\authornote{Corresponding author.}
\email{yuanzhou@zstu.edu.cn}
\orcid{0000-0002-1583-7570}
\affiliation{%
  \institution{Zhejiang Sci-Tech University, China}
  \country{}
}





\renewcommand{\shortauthors}{Mingfei Cheng et al.}

\begin{abstract}
Search-based testing is critical for evaluating the safety and reliability of autonomous driving systems (ADSs). 
However, existing approaches are often built on heterogeneous frameworks (e.g., distinct scenario spaces, simulators, and ADSs), which require considerable effort to reuse and adapt across different settings. 
To address these challenges, we present \tool{}, a unified and extensible infrastructure for search-based ADS testing built on the widely used CARLA simulator. 
\tool{} introduces a unified scenario definition, \textit{OpenScenario}, that specifies scenarios using low-level, actionable parameters to ensure compatibility with existing methods while supporting extensibility to new testing designs (e.g., multi-autonomous-vehicle testing).
On top of this, \tool{} decouples the \textit{testing engine}, \textit{scenario execution}, and \textit{ADS integration}. 
The \textit{testing engine} leverages evolutionary computation to explore new scenarios and supports flexible customization of core components. 
The \textit{scenario execution} can run arbitrary scenarios using a parallel execution mechanism that maximizes hardware utilization for large-scale batch simulation.
For \textit{ADS integration}, \tool{} provides access to 12 ADSs through a unified interface, streamlining configuration and simplifying the incorporation of new ADSs. 
Our tools are publicly available at \href{https://github.com/MingfeiCheng/Drivora}{https://github.com/MingfeiCheng/Drivora}.  
\end{abstract}




\maketitle

\section{Introduction}
\newcommand{\cmark}{\ding{51}} 
\newcommand{\xmark}{\ding{55}} 

Autonomous Driving Systems (ADSs) are a groundbreaking technology that enables vehicles to operate without human intervention. 
They rely on a combination of sensors and artificial intelligence algorithms to perceive the environment, make decisions, and navigate safely. 
Ensuring the safety and reliability of ADSs before deployment requires thorough testing.

Simulation-based testing, particularly search-based approaches~\cite{av_fuzzer, cheng2023behavexplor, icse_samota}, has become a mainstream method. 
However, many state-of-the-art ADS testing techniques~\cite{av_fuzzer, cheng2023behavexplor, icse_samota} are implemented on heterogeneous frameworks (e.g., distinct scenario spaces, simulators, and ADSs), which makes reuse and adaptation across different settings challenging. 
For example, AVFuzzer~\cite{av_fuzzer} and BehAVExplor~\cite{cheng2023behavexplor} are both built on LGSVL~\cite{LGSVLSunsetting} with different scenario configurations; with the deprecation of LGSVL~\cite{LGSVLSunsetting}, their reusability and migration to other platforms become problematic. 
These limitations underscore the need for an open-source, unified, and extensible infrastructure for ADS testing that minimizes integration effort. 



CARLA~\cite{dosovitskiy2017carla} has become the most widely adopted simulator for ADS evaluation, and recent studies~\cite{carla_leaderboard, jia2024bench, liu2025towards, tehrani2025pcla} have attempted to build unified evaluation platforms on top of it. 
Nevertheless, as shown in Table~\ref{tab:diff-platform}, existing CARLA-based evaluation platforms exhibit two main limitations:  
\ding{182} \textbf{Static scenario definition.} Current platforms primarily test ADSs using manually crafted static scripts, which limits support for advanced testing methods.  
\ding{183} \textbf{Inefficient scenario execution.} Existing execution is restricted to a single-AV setting (i.e., only one AV controlled by the ADS under test) and remains sequential (i.e., running only one scenario at a time). This limitation constrains the testing space by hindering extensibility to diverse settings such as multi-AV scenarios, and it also limits testing efficiency due to underutilized hardware resources.


\begin{table}[!t]
    \centering
    \caption{Comparison with ADS evaluation frameworks.}
    \small
    \vspace{-5pt}
    \resizebox{1.0\linewidth}{!}{
    \begin{tabular}{l|cc|cc}
    \toprule
    \multirow{2.5}*{\textbf{Platform}} 
        & \multicolumn{2}{c|}{\textbf{Scenario Definition}} 
        & \multicolumn{2}{c}{\textbf{Scenario Execution}} \\
    \cmidrule(lr){2-3} \cmidrule(lr){4-5}
        & \textbf{Static Script} 
        & \textbf{Support Testing}
        & \textbf{AV Scalability}  
        & \textbf{Parallel} \\
    \midrule
    Leaderboard~\cite{carla_leaderboard}   & \cmark & \xmark & \xmark & \xmark \\
    Bench2Drive~\cite{jia2024bench} & \cmark & \xmark & \xmark & \xmark \\
    V2Xverse~\cite{liu2025towards} & \cmark & \xmark & \xmark & \xmark \\
    PCLA~\cite{tehrani2025pcla} & \cmark & \xmark & \xmark & \xmark \\
    \midrule
    \tool{} & \cmark & \cmark & \cmark & \cmark \\
    \bottomrule
    \end{tabular}
    }
    \vspace{-5pt}
    \label{tab:diff-platform}
\end{table}

To overcome these limitations, we present \tool{}, which integrates different ADSs and testing techniques
into a unified infrastructure. \tool{} first introduces a unified scenario definition, \textit{OpenScenario}, which defines scenarios with low-level actionable parameters (e.g., waypoints), ensuring compatibility with existing techniques while supporting extensibility to new testing designs. 
To further enhance extensibility, \tool{} decouples the testing pipeline into three components: \textit{testing algorithm}, \textit{scenario execution}, and \textit{ADS integration}, facilitating the incorporation of different testing methods and ADSs. 
The testing framework in \tool{} is built on a classical evolutionary computation paradigm, enabling rapid prototyping and evaluation of different testing methods. 
For \textit{scenario execution}, \tool{} provides a unified executor that handles arbitrary scenarios with any number of AVs, thereby supporting multi-AV settings. 
It further incorporates a parallel execution mechanism that enables batch-wise simulation (e.g., running four individuals with two executors), a capability particularly advantageous for search-based approaches.
For \textit{ADS integration}, \tool{} adopts a unified agent interface that ensures compatibility across existing ADSs and simplifies the extension to new ones. 
Currently, \tool{} integrates 12 ADSs and five state-of-the-art testing methods, all of which are available for direct use. 
This design substantially reduces the complexity and time required to set up simulation environments, allowing researchers, especially from the software engineering community, to focus on testing method evaluation and innovation rather than environment configuration.

\section{OpenScenario Definition}

\begin{table}[!t]
\centering
\caption{OpenScenario Template.}
\small
\vspace{-5pt}
\resizebox{1.0\linewidth}{!}{
\begin{tabular}{l|l|p{0.45\linewidth}}
\toprule
\textbf{Entities} & \textbf{Entity Attributes} & \textbf{Description} \\
\midrule
\multirow{5}{*}{\texttt{ego\_vehicles}} 
  & id           & Unique identifier of the vehicle \\
  & model        & Vehicle type \\
  & route        & Predefined trajectory of the AV \\
  & start\_time  & Start time of the AV \\
  & config\_path & Config of the ADS under test \\
\midrule
\multirow{4}{*}{\texttt{npc\_vehicles}} 
  & id           & Unique identifier of the vehicle \\
  & model        & Vehicle type \\
  & route        & Predefined trajectory \\
  & start\_time  & Start time of the vehicle \\
\midrule
\multirow{4}{*}{\texttt{npc\_walkers}} 
  & id           & Unique identifier of the walker \\
  & model        & Pedestrian type \\
  & route        & Predefined walking path \\
  & start\_time  & Start time of the walker \\
\midrule
\multirow{3}{*}{\texttt{npc\_obstacles}} 
  & id           & Unique identifier of the obstacle \\
  & model        & Obstacle type \\
  & location     & Static position on the map \\
\midrule
\multirow{2}{*}{\texttt{map\_region}} 
  & town         & The town name (e.g., Town01) \\
  & region\_segment & Specific region in the map \\
\midrule
\multirow{1}{*}{\texttt{weather}} 
  & parameters   & Weather parameters in CARLA \\
\midrule
\multirow{3}{*}{\texttt{traffic\_lights}} 
  & green\_time  & Duration of green light \\
  & red\_time    & Duration of red light \\
  & yellow\_time & Duration of yellow light \\
\bottomrule
\end{tabular}
}
\vspace{-5pt}
\label{tab:scenario-dsl}
\end{table}

Scenarios are fundamental elements of ADS testing, typically characterized by a set of parameters that specify the environment (e.g., weather conditions) and traffic participants such as Non-Player Character (NPC) vehicles. 
The primary objective of ADS testing is to automatically generate scenarios that evaluate system performance under different requirements, such as reaching a destination without encountering safety-critical events. 
Different testing methods~\cite{av_fuzzer, cheng2023behavexplor, icse_samota, huai2023doppelganger, zhou2023specification, lu2022learning, css_drivefuzzer, cheng2025decictor} often define scenarios with heterogeneous attribute spaces, but they can ultimately be expressed using low-level actionable parameters. 
For example, a lane-changing NPC vehicle can be represented as a sequence of waypoints that define its trajectory, which the vehicle then follows. 

Therefore, to ensure compatibility across different testing techniques, \tool{} provides an \textit{OpenScenario} template, as shown in Table~\ref{tab:scenario-dsl}. It specifies the following entities using low-level parameters: 
(1) \texttt{ego\_vehicles} specifies a list of AV controlled by the ADS under test, including a unique identifier, vehicle type (e.g., \texttt{lincoln.mkz2017}), route (i.e., start position and destination), start time, as well as 
the configuration path of the ADS.  
(2) \texttt{npc\_vehicles}, \texttt{npc\_walkers}, and  \texttt{npc\_obstacles} represent lists of dynamic and static actors in the environment, each defined by a unique identifier, model, route or location (e.g., a list of predefined waypoints controlling their motion), and start time.  
(3) \texttt{map\_region} defines the simulation town and the specific region segment (e.g., \texttt{min\_x}, \texttt{max\_x}) in which the scenario is executed.  
(4) \texttt{weather} encapsulates CARLA’s \texttt{WeatherParameters}, including cloudiness, precipitation, wind, fog, and sun conditions.  
(5) \texttt{traffic\_lights} sets the timing of green, yellow, and red lights. 
The design of \textit{OpenScenario} can be extended with additional entities and attributes to accommodate specific research needs. 

\section{Tool Design}


\begin{figure*}[!t]
    \centering
    \includegraphics[width=0.8\linewidth]{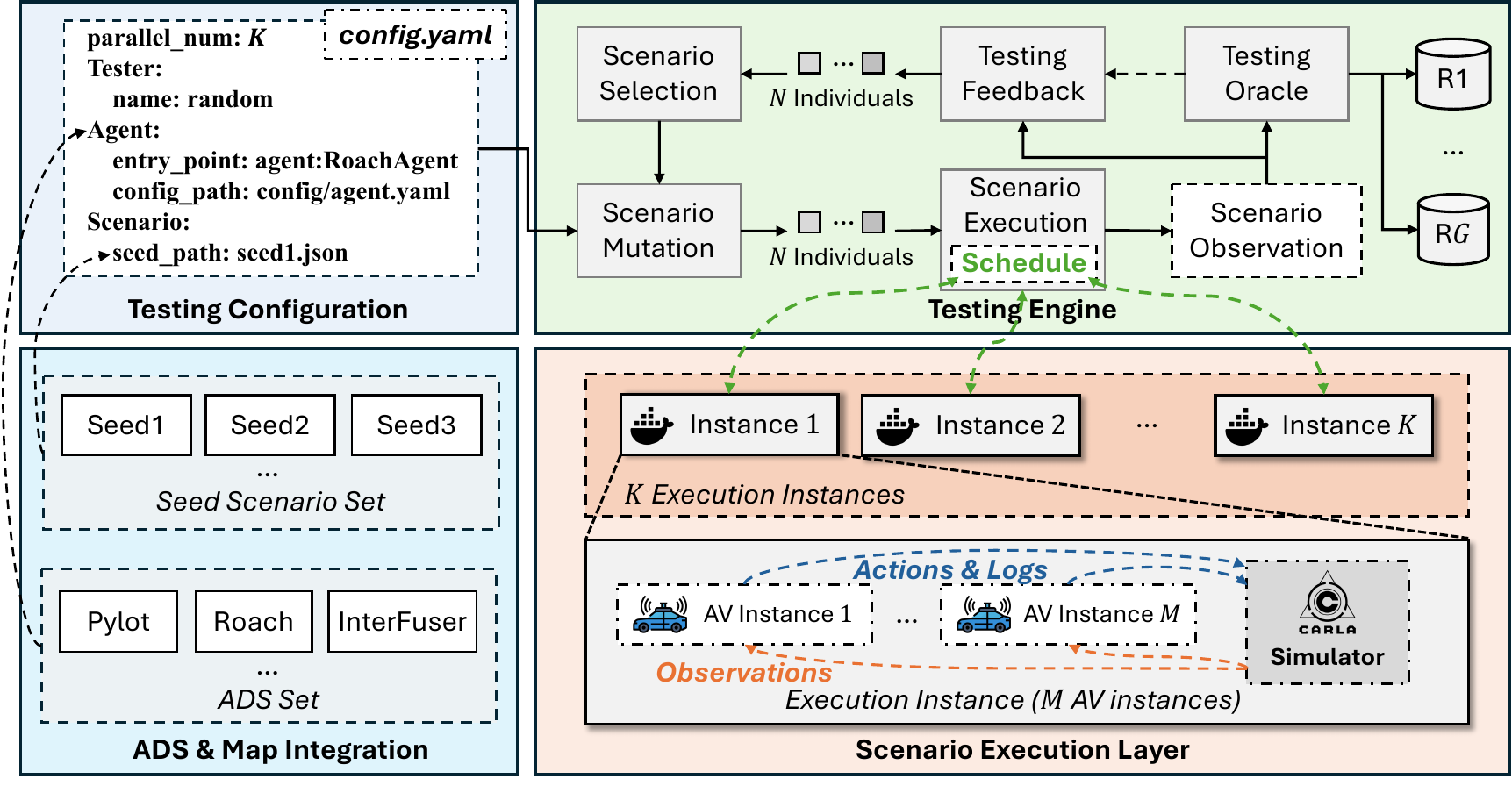}
    \vspace{-5pt}
    \caption{The overall architecture of the \tool{} framework.}
    \vspace{-5pt}
    \label{fig:tool-framework}
\end{figure*}

Figure~\ref{fig:tool-framework} depicts the architecture of \tool{}, including four main components: 
ADS \& Seed Integration, 
Testing Configuration, 
Testing Engine and Parallel Execution Layer.

\textbf{ADS \& Seed Integration.}  
This component provides the initial resources for testing, including ADSs under test and seed scenarios defined by \textit{OpenScenario}.  
(1) \textit{ADS Set} provides a collection of ADSs under test. To simplify integration and ensure consistency across different systems, \tool{} offers a unified agent interface that exposes two core APIs: 
\texttt{setup\_env()}, which initializes the simulator, configures controlled actors, and sets up the ADS agent; and 
\texttt{run\_step()}, which advances the simulation by one step, returning ADS control actions and recording the agent’s internal logs into the scenario observation for analysis and debugging.
(2) \textit{Seed Scenario Set} contains seed scenarios that define the ego task through the \texttt{ego\_vehicles} route and the \texttt{map\_region} entities in Table~\ref{tab:scenario-dsl}. 

\textbf{Testing Configuration.} This component defines the configurations required by \tool{}, including:  
\textit{Tester}, which specifies the testing algorithm;  
\textit{Agent}, which selects the ADS under test from the ADS Set; and  
\textit{Scenario}, which defines the initial seed for testing.  

\textbf{Testing Engine.}  
As shown in Figure~\ref{fig:tool-framework}, the \textit{Testing Engine} constitutes the core of \tool{}, enabling search-based algorithms to efficiently identify critical scenarios (e.g., collisions). 
It consists of five components:  
(1) \textit{Scenario Mutation} generates a population of $N$ individual scenarios by mutating attributes defined in Table~\ref{tab:scenario-dsl};  
(2) \textit{Scenario Execution} evaluates the $N$ scenarios in the simulator. 
To improve efficiency, the \textit{Scenario Execution Layer} schedules these individuals across $K$ parallel execution instances;  
(3) \textit{Testing Oracle} verifies whether the ADS satisfies $G$ predefined testing requirements (e.g., detecting collisions);  
(4) \textit{Testing Feedback} provides guidance based on testing objectives (e.g., using the minimum distance to NPC vehicles when searching for collisions);  
(5) \textit{Scenario Selection} chooses promising candidates for further mutation and evaluation.  
Each component in \textit{Test Engine} can be customized and extended to support different testing requirements.


\textbf{Scenario Execution Layer.}  
As shown in Figure~\ref{fig:tool-framework}, this layer supports scenarios with multiple AVs ($M \geq 1$) and parallel execution ($K \geq 1$), thereby maximizing hardware utilization and improving testing efficiency. 
Each execution instance $k \in \{1,\dots,K\}$ runs in an isolated container, accepts a scenario defined according to the template in Table~\ref{tab:scenario-dsl}, and produces a sequence of \texttt{scene\_obs}. 
Within an execution instance, each AV $m \in \{1,\dots,M\}$ controlled by the ADS under test interacts independently with the shared simulator. 
As summarized in Table~\ref{tab:scene-obs}, each \texttt{scene\_obs} records runtime information (e.g., simulation step, timestamp, actor states, and agent logs), which is used to compute testing feedback and oracle outcomes.



\begin{table}[!t]
\centering
\caption{Structure of a \texttt{scene\_obs}.}
\vspace{-5pt}
\small
\resizebox{\linewidth}{!}{
\begin{tabular}{l|p{0.6\linewidth}}
\toprule
\textbf{Key} & \textbf{Description} \\
\midrule
\texttt{step}        & Simulation step index $i$ \\
\texttt{timestamp}   & Simulation time $t$ at this step \\
\texttt{egos}        & States of ego vehicles \\
\texttt{other\_actors} & States of NPC actors and traffic lights \\
\texttt{agent\_logs} & Internal logs of ADS agents \\
\bottomrule
\end{tabular}
}
\vspace{-10pt}
\label{tab:scene-obs}
\end{table}

\section{Usage}  
\tool{} runs on Linux, with all core components implemented in Python and simulation supported by a Docker-based CARLA backend.  
We provide detailed installation guidelines, including CARLA and Docker setup~\cite{ourweb}.  

\textbf{Integrated Toolkit.} \tool{} has integrated \textbf{12 ADSs}, including module-based, end-to-end, and vision–language-based systems, and \textbf{5} state-of-the-art testing techniques. Details are available at~\cite{ourweb}.
 
\textbf{Testing Environment Setup.} 
ADSs and testing techniques often rely on heterogeneous libraries, which can easily lead to dependency conflicts.  
To address this, \tool{} provides configuration scripts that automatically set up the required environment for each ADS and testing method.  
We recommend managing separate Python environments (e.g., Anaconda or \texttt{venv}) to ensure isolation.  
For example, to test ADS \texttt{ads\_name} with testing method \texttt{tester\_name} under CARLA version \texttt{carla\_ver} (the version used in the ADS), users simply run:  
\begin{tcolorbox}[colback=black!5!white,colframe=black!40,boxrule=0.2pt,
left=2pt,right=2pt,top=1pt,bottom=1pt]
\begin{verbatim}
bash install.sh [ads_name] [tester_name] [carla_ver]
\end{verbatim}
\end{tcolorbox}


\textbf{Seed Generation.}  
While human-crafted seed scenarios are acceptable, \tool{} also provides scripts to automatically generate customized ones, such as ego tasks with varying trajectory lengths. 
Users can specify parameters, such as the number of seeds, target town, and route constraints.  
For example, the following command generates 10 seed scenarios in \texttt{Town01} with route lengths between 50 and 200 meters, and saves them under \texttt{scenario\_datasets}:  
\begin{tcolorbox}[colback=black!5!white,colframe=black!40,boxrule=0.2pt,
left=2pt,right=2pt,top=1pt,bottom=1pt]
\begin{verbatim}
python -m seed_generator.open_scenario --num 10 \
    --town Town01 --min_length 50 --max_length 200 
\end{verbatim}
\end{tcolorbox}

\textbf{Run Testing.}  
The main configuration is specified in \texttt{config.yaml} (key attributes are shown in Figure~\ref{fig:tool-framework}) and can be customized by users. 
In addition, we provide example scripts in the \texttt{scripts} directory for direct use, such as testing ADS \texttt{ads\_name} with testing method \texttt{tester\_name}:  
\begin{tcolorbox}[colback=black!5!white,colframe=black!40,boxrule=0.2pt,
left=2pt,right=2pt,top=1pt,bottom=1pt]
\begin{verbatim}
bash scripts/[tester_name]/[ads_name].sh
\end{verbatim}
\end{tcolorbox}

\textbf{Output Format.} For each test run, \tool{} automatically saves logs, scenario configurations, and execution observations for further analysis and debugging. Output structure is available at~\cite{ourweb}.

\section{Demo Results}
\begin{figure*}
    \centering
    \includegraphics[width=0.85\linewidth]{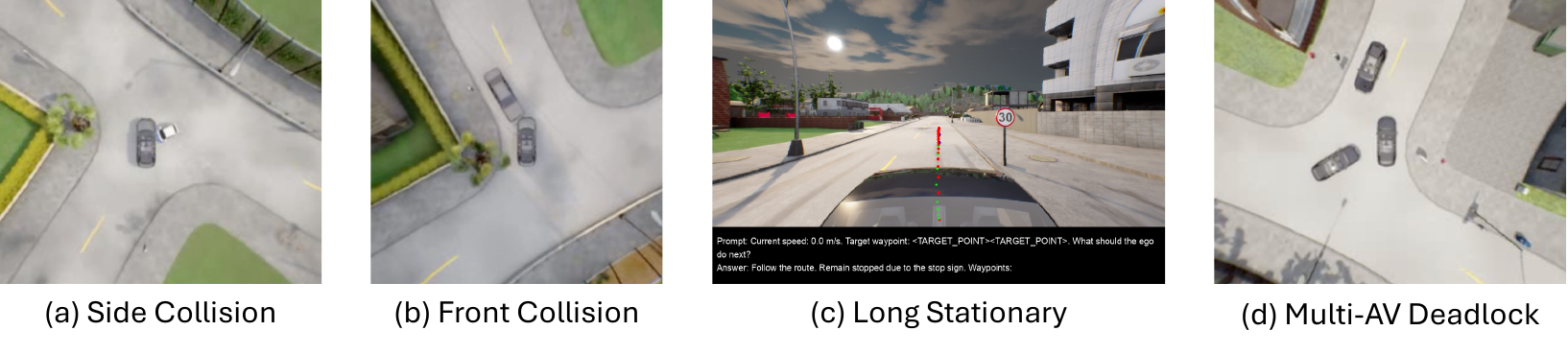}
    \vspace{-10pt}
    \caption{Illustration of violation cases.}
    \vspace{-5pt}
    \label{fig:demo-cases}
\end{figure*}


\textbf{Violation Discovery.}  
Figure~\ref{fig:demo-cases}(a)–(c) presents three representative violation cases discovered using \tool{}. 
Figure~\ref{fig:demo-cases}(a) shows the ADS accelerating competitively with other vehicles, resulting in a side collision at a junction. 
Figure~\ref{fig:demo-cases}(b) illustrates a failure to make a correct deceleration decision when facing an NPC vehicle, leading to a front collision. 
Figure~\ref{fig:demo-cases}(c) depicts a prolonged stationary case in a vision–language-based ADS, caused by misinterpreting a speed-limit sign as a stop sign. 
Together, these cases demonstrate that \tool{} can effectively uncover safety-critical violations and reveal the ADS under test's limited safety capability. 
Here, we only present demo results to illustrate the usefulness of \tool{}; we plan to conduct systematic experiments across different ADSs and testing methods to provide more comprehensive empirical insights.

\noindent \textbf{Execution Efficiency.}  
We evaluate execution efficiency by measuring the number of scenarios executed within a fixed time budget, varying the number of individuals ($N$) and parallel execution instances ($K$), where $N \geq K$. 
We find that throughput grows almost linearly with the number of parallel instances ($K$), validating the design of the parallel execution layer. 
This scalability enables efficient large-scale testing, making it practical to explore broader scenario spaces within limited time budgets.  

\noindent \textbf{Multi-AV Testing.}  
Another key capability of \tool{} is its support for multi-AV testing, where the same ADS controls multiple AVs. 
As shown in Figure~\ref{fig:demo-cases}(d), \tool{} identifies a deadlock scenario involving three AVs, an emergent behavior that cannot be captured in single-AV settings. 
Thus, \tool{} not only facilitates testing of ADS performance in single-vehicle tasks but also extends to traffic-level evaluations, where interactions among multiple AVs expose new classes of safety risks.


\section{Conclusion \& Future Work}
In this paper, we presented \tool{}, a unified and extensible infrastructure for search-based testing of autonomous driving systems (ADSs). 
\tool{} introduces a low-level actionable scenario definition, \textit{OpenScenario}, which unifies existing testing methods and supports extensibility to diverse settings, such as multi-AV testing. 
It further adopts a decoupled architecture consisting of the \textit{testing engine}, \textit{scenario execution}, and \textit{ADS integration}. 
The modular testing engine enables rapid prototyping of new algorithms, while the scenario execution layer supports both multi-AV testing and parallel execution to maximize hardware utilization. 
For convenient experimentation, \tool{} integrates 12 ADSs via a unified API.
Looking ahead, we plan to continuously maintain and extend \tool{}. Future directions include conducting comprehensive evaluations of diverse ADSs with different testing techniques to derive empirical insights, as well as incorporating advanced scenario mutations, such as reinforcement learning–based approaches.

\section*{Acknowledgments}
This work was supported by the Natural Science Foundation of China under Grant 62572441.
Lionel Briand was supported by the Canada Research Chair and Discovery Grant programs of the Natural Sciences and Engineering Research Council of Canada (NSERC) and the Research Ireland grant 13/RC/2094-2.




\bibliographystyle{ACM-Reference-Format}
\bibliography{main}
\end{document}